\documentstyle[preprint,aps,epsf,eqsecnum]{revtex}
\tightenlines
\begin{document}
\def\non{\nonumber}
\def\be{\begin{eqnarray}}
\def\en{\end{eqnarray}}
\def\la{\langle}
\def\ra{\rangle}
\def\hep{\hat{\varepsilon}}
\def\pr{Phys. Rev.~}
\def\prl{Phys. Rev. Lett.~}
\def\pl{Phys. Lett.~}
\def\np{Nucl. Phys.~}
\def\zp{Z. Phys.~}
\def\bi{\bibitem}
\pagestyle{empty}                                      
\draft
\vfill
\title{Lepton pair decays of the $K_L$ meson in the light-front model}

\author{C.~Q.~Geng$^{a,b}$ and C.~W.~Hwang$^a$}
\address{\sl $^a$ Department of Physics, National Tsing Hua University,
Hsinchu, Taiwan 300, ROC}
\address{\sl $^b$ Theory Group, TRIUMF, 4004 Wesbrook Mall,
Vancouver, B.C. V6T 2A3, Canada}
%
%
\vfill
\maketitle
\begin{abstract}
We analyze $K_L$ lepton pair decays of $K_L \to l^+\l^-\gamma$ and
$K_L \to l^+l^-l'^+l'^-$ $(l,\,l'=e,\, \mu)$ within the framework
of the light-front QCD approach (LFQA). With the $K_L \to \gamma^* 
\gamma^*$ form factors evaluated in a model with the LFQA, we calculate 
the decay branching ratios and find out that our results
 are all consistent with the experimental data.
In addition, we study  $K_L \to l^+l^-$ decays. We point out
that our prediction on $K_L\to e^+e^-$ is
about $20\%$ smaller
than that in the ChPT. We also discuss whether one could extract the
short-distance physics from $K_L\to \mu^+\mu^-$.
\end{abstract}
%
%
\pagestyle{plain}
\section{Introduction}
The study of kaon decays has played a pivotal role in formulating the
standard model of electroweak interactions \cite{Kaon}.
In particular, the rare
decay of $K_L\to\mu^+\mu^-$ was used to constrain the flavor changing
neutral current \cite{FCNC} as well as the top quark mass \cite{Geng}.
However, there are ambiguities in extracting the short-distance
contribution since the long-distance
contribution dominated by the two-photon intermediate state is not well
known because its dispersive part cannot be calculated
in a reliable way \cite{Kmm,Valencia,Kmm1,Kmm2}.
To have a better understanding of this
dispersive part, it is important to study the
 lepton pair decays of the $K_L$ meson such as $K_L\to l^+l^-\gamma$ and
$l^+l^-l^+l^-$ ($l=e,\mu$) since they can provide us with information
on the structure of the $K_L\to\gamma^*\gamma^*$ vertex
\cite{Kmm,Valencia,Kmm1,Kmm2}.
On the other hand, since these lepton pair decays are dominated by the
long-distance physics, they can also be served
as a testing ground for theoretical techniques such as
 chiral Lagrangian or other non-perturbative methods
that seek to account for the low-energy behavior of QCD.

Recently, several new measurements of
the decay branching ratios of  $K_L \to \mu^+\mu^- \gamma$,
$K_L \to e^+e^-e^+e^-$, and $K_L \to e^+e^-\mu^+\mu^-$ have
been reported \cite{prl2m,prl4e,NA48,prl2e2m}. These decays proceed
entirely through the $K\gamma^*\gamma^*$ vertex and provide
the best opportunity for the study of its form factor.
In Ref. \cite{MT}, since the assumption of neglecting the momentum
dependence for the form factor was adopted,  the
results for the decays are only valid for those with only
the electron-positron pair.
In Ref. \cite{ZG}, the decays were studied at the order
 $p^6$ in Chiral Perturbation Theory (ChPT).
However, all the results in Ref. \cite{ZG} are smaller than the current
experimental values.
 In this work, we consider another non-perturbative method in
the LFQA to analyze the
$K\gamma^*\gamma^*$ form factor. As is well known \cite{LFQM},
the LFQA  allows an exact
separation in momentum space between the center-of-mass motion
and intrinsic wave functions. A consistent treatment of quark
spins and the center-of-mass motion can also be carried out.
It has been successfully applied  to calculate various form
factors \cite{LFQMa,Dem,Cheung2,cheng}.

The paper is organized as follows. In Sec. II, we derive the theoretical
 formalism for the decay
constant and the $K\gamma^*\gamma^*$ vertex and use these
formalism in the LFQA  to extract the decay constant and the form
factor. In Sec. III, we fix the parameters appearing in the wave
functions and calculate the form factors and branching ratios.
Finally, conclusions are given in Sec. IV.
\section{Framework}
We start with the
 $K$ meson decay constant $f_K$,
 defined by
\be
\la 0|A^\mu|K(P)\ra=\,if_K P^\mu,\label{decaydef}
\en
where $A^\mu=\bar u \gamma^\mu \gamma_5 s$ is the
axial vector current. Assuming a constant vertex
function $\Lambda_K$ \cite{Dem,Jaus} which is related
to the $u\bar s$ bound state of the kaon.
Then the quark-meson diagram, depicted in Fig.~1 (a), yields
\be
\la 0|A^\mu|K(P)\ra=-\sqrt{N_c}\int {d^4p_1\over{(2\pi)^4}}
\Lambda_K\text{Tr}\Bigg[\gamma_5{i(\not{\!p_2}+m_s)
\over{p_2^2-m^2_s+i\epsilon}}\gamma^\mu\gamma_5{i(\not{\!p_1}+m_u)
\over{p_1^2-m^2_u+i\epsilon}}\Bigg], \label{decay1}
\en
where $m_{u,s}$ are the masses of $u$ and $s$ quark, respectively,
 and $N_c$ is the number of colors. We consider the poles in
denominators in terms of the LF coordinates $(p^-,p^+,p_\perp)$
and perform the integration over the LF ``energy" $p_1^-$ in Eq.
(\ref{amp}). The result is
\be
\la 0|A^\mu|K(P)\ra = \sqrt{N_c}\int {[d^3 p_1]\over{p^+_1p^+_2}}
           {\Lambda_{K_L} \over{P^--p^-_{1{\rm on }}-p^-_{2{\rm on }}}}
(I^\mu_1|_{p^-_1=p^-_{1{\rm on}}}), \label{decay2}
\en
where
\be
[d^3 p_1]&=&{dp^+_1 d^2p_{1\perp}\over{2(2\pi)^3}},~~~p^-_{i{\rm on}}
={m^2_i+p^2_{i\perp}\over{p^+_i}}, \non \\
I^\mu_1&=&{\rm Tr}[\gamma_5(\not{\!
p_2}+m_s)\gamma^\mu\gamma_5(\not{\! p_1}+m_u)].
\en
For $K_L\to \gamma^*\gamma^*$,
with the assumption of $CP$ conservation the amplitude is given 
by 
\be
A(K_L\to \gamma^*(q_1,\epsilon_1)~\gamma^*(q_2,\epsilon_2))
=iF(q^2_1,q^2_2)~\varepsilon_{\mu\nu\rho\sigma}~\epsilon^\mu_1
 ~\epsilon^\nu_2 ~q^\rho_1 ~q^\sigma_2, \label{def}
\en
where the form factor of $F(q^2_1,q^2_2)$ in Eq. (\ref{def})
is a symmetric function under the interchange of $q^2_1$ and $q^2_2$.
In our model, by using the same procedure
 as above, from the quark-meson diagram depicted in Fig.~2 
we get
\be
A(K_L\to \gamma^*~\gamma^*)&=&-\int {d^4 p_1 \over{(2
\pi)^4}} \Lambda_{K_L}\Bigg\{{\rm Tr}\Bigg[\gamma_5
        {i(\not{\! p_3}+m_s)\over{p_3^2-m^2_s+i\epsilon}}\not{\! \epsilon_2}
       {i(\not{\!p_2}+m_s)\over{p_2^2-m^2_s+i\epsilon}} \non \\
&&~~~~~~~~~~~~~~~~~~\times C_W(q^2_1)\not{\! \epsilon_1}
        {i(\not{\! p_1}+m_d)\over{p_1^2-m^2_d+i\epsilon}}
+(d\leftrightarrow s)\Bigg]+(\epsilon_1 \leftrightarrow \epsilon_2) \Bigg\},
        \label{amp}
\en
where $p_2=p_1-q_1$, $p_3=p_1-P$, and $C_W$ is the effective
contribution to the inclusive $s\to d \gamma^*$ decay. After
integrating over $p_1^-$,
 we obtain
\be
A(K_L\to \gamma^*~\gamma^*) &=&\Bigg\{\Bigg[\int^{q_1}_0 {[d^3 p_1]
\over{\prod^3_{i=1} p^+_i}}
           {\Lambda_{K_L} \over{P^--p^-_{1{\rm on }}-p^-_{3{\rm on }}}}
(I_2|_{p^-_1=p^-_{1{\rm on}}})
           {C_W(q^2_1)\over{q^-_1-p^-_{1{\rm on }}-p^-_{2{\rm on }}}}
           \non \\
&&~+\int^P_{q_1}  {[d^3 p_1]\over{\prod^3_{i=1} p^+_i}}
          {\Lambda_{K_L} \over{P^--p^-_{1{\rm on }}-p^-_{3{\rm on }}}}
(I_2|_{p^-_3=p^-_{3{\rm on}}})
            {C_W(q^2_1) \over{q^-_2-p^-_{2{\rm on }}-p^-_{3{\rm on }}}}
+(d\leftrightarrow s)\Bigg]\non \\
&&~+(\epsilon_1 \leftrightarrow \epsilon_2)\Big\},
            \label{pole}
\en
where $q^-_2=P^--q^-_1$ and
\be
I_2&=&{\rm Tr}[\gamma_5(\not{\!
p_3}+m_s)\not{\! \epsilon_2}(\not{\! p_2}+m_s)\not{\! \epsilon_1}
(\not{\! p_1}+m_d)].
\en
We note that
we do not expect that the absolute decay 
widths of $K_L \to l^+l^-\gamma$ and $K_L\to\gamma \gamma$  
calculated from Eq. (\ref{pole}) can fit to the experimental values 
\cite{sdg}. 
However, we can estimate the relative form factors of 
these leptonic decays versus the two-photon decay, and compare the 
 branching ratios with the experimental ones. 
Recent works on both short-distance (SD) and long-distance (LD)
contributions to $s \to d\gamma^*$ can be found in Ref. 
\cite{He}.

As described in Ref. \cite{hwcw1}, the vertex function
$\Lambda_{K_L}$
 and the denominators in Eq. (\ref{pole}) correspond to the $K_L$
meson bound state. In the LFQA, the internal structure of the
meson bound state \cite{Cheung2,cheng,hwcw2} consists of $\phi$, which
describes the momentum distribution of the constituents in the
bound state, and $R^{S,S_z}_{\lambda_1,\lambda_2}$, which creates a
state of definite spin ($S,S_z$) out of LF helicity
($\lambda_1,\lambda_2$) eigenstates and is related to the Melosh
transformation \cite{Melosh}. A convenient approach
relating these two parts is shown in Ref. \cite{hwcw1}. The interaction
Hamiltonian is assumed
to be $H_I=i\int d^3 x \bar {\Psi}\gamma_5 \Psi \Phi $ where
$\Psi$ is the quark field and $\Phi$ is the meson field containing $\phi$
and $R^{S,S_z}_{\lambda_1,\lambda_2}$. When considering the
normalization of the meson state depicted in Fig.~1 (b)
in the LFQA, we obtain
\be
\la M (P',S',S'_z)|H_I~H_I|M(P,S,S_z)\ra
&=&2(2\pi)^3\delta^3(P'-P)\delta_{SS'}\delta_{S_zS'_z}
\non \\
&\times& \int [d^3p_1]\phi^2 R^{S,S_z}_{\lambda_1,\lambda_2}
R^{S',S'_z}_{\lambda_1,\lambda_2} \text{Tr}\Bigg[\gamma_5
{-\not{\!p_2}+m_2\over{p_2^+}}\gamma_5{\not{\!p_1}+m_1\over{p_1^+}}\Bigg].\non \\
\en
If we normalize the meson state and the momentum distribution function
$\phi$ as \cite{Cheung2}
\be
\la M (P',S',S'_z)|H_I~H_I|M(P,S,S_z)\ra
=2(2\pi)^3P^+\delta^3(P'-P)\delta_{SS'}\delta_{S_zS'_z},
\en
and
\be
\int {d^3p_1 \over{2(2\pi)^3}} {1\over{P^+}} |\phi|^2=1,
\en
respectively, where $p_1$ and $p_2$ are the on-mass-shell momenta,
we have that
\be
R^{S,S_z}_{\lambda_1,\lambda_2}={\sqrt{p^+_1 p^+_2}
                  \over{2\sqrt{p_{1{\rm on}}\cdot p_{2{\rm on}}+m_1 m_2}}}.
\en
The wave function and the Melosh transformation of the meson are related
to the bound state vertex function $\Lambda_M$ by
\be
{\Lambda_M \over{P^--p^-_{1{\rm on }}-p^-_{2{\rm on }}}}
\longrightarrow R^{S,S_z}_{\lambda_1,\lambda_2}~\phi_M\,.
\en
We note that $p_1$, $p_2$  and $p_3$ in the trace of $I_{1,2}$ must be on
the mass shell for self-consistency.
 After taking the ``good " component $\mu=+$, we use the
definitions of the LF momentum variables
$(x,x',k_\perp,k'_\perp)$ \cite{cheng} and take a Lorentz frame
where $P_\perp=P'_\perp=0$  to have $q_\perp=0$ and
$k'_\perp=k_\perp$. The decay constant $f_K$ and the form
factor $F(q_1^2,q^2_2)$ can be extracted by comparing these
results with Eqs. (\ref{decaydef}) and (\ref{def}), respectively,
$i.e.$,
\be
f_K=\,2\sqrt{2}\sqrt{N_c}\int {dx\,d^2k_\perp\over 2(2\pi)^3}\,{\phi_{K_L}(x,
k_\perp)\over\sqrt{a^2+k_\perp^2}}\,a\,, \label{fp}
\en
and
\be
F(q_1^2,q^2_2)&=& \int {d^2k_\perp\over 2(2\pi)^3}
\Bigg\{C_W(q^2_1)\Bigg[\int^{r_+}_0dx~ {\phi_{K_L}
(x,k_\perp)\over\sqrt{a^2 +k_\perp^2}}{a[r_+/(r_+-x)]
\over{{m^2_s+k^2_\perp\over{(x/r_+)}}
+{m^2_s+k^2_\perp\over{1-(x/r_+)}}-q^2_2}}\non \\
&&~~~~~~~~~~~~~~~~~~~~~+\int^{1}_{r_+}dx~
{\phi_{K_L}(x,k_\perp)\over\sqrt{a^2 +k_
\perp^2}}{a[(1-r_+)/(x-r_+)]\over{{m^2_d+k^2_\perp
\over{(1-x)/(1-r_+)}}+{m^2_s+k^2_\perp\over{(x-r_+)/(1-r_+)}}-q^2_1}}
+(d\leftrightarrow s)\Bigg]\non \\
&&~~~~~~~~~~~~+(q_1 \leftrightarrow q_2;r_+ \leftrightarrow 1-r_-) \Bigg\},
 \label{Hi}
\en
where
\be
a&=&m_{u,d}x+m_s(1-x)\,,\ \ m_u\;=\;m_d\,,\non \\
r_\pm&=&{1\over {2M^2_{K_L}}}\big[M_{K_L}^2+q^2_1-q^2_2\pm
\sqrt{(M^2_{K_L}
+q^2_1-q^2_2)^2-4 M^2_{K_L} q^2_1}~\big], \label{y12}
\en
and $x$ is
the momentum fraction carried by the spectator antiquark in the
initial state.

In principle, the momentum distribution amplitude $\phi(x,k_\bot)$
can be obtained by solving the LF QCD bound state
equation\cite{Zhang}. However, before such first-principle
solutions are available, we shall have to use phenomenological
amplitudes. One momentum distribution function that has often
 been used in the literature for mesons is the Gaussian-type,
\begin{equation}
    \phi(x,k_\perp)_{\rm G}={\cal N} \sqrt{{dk_z\over dx}}
        ~{\rm exp}\left(-{\vec k^2\over 2\omega^2}\right),
        \label{gauss}
\end{equation}
where ${\cal N}=4(\pi/\omega^2)^{3/4}$ and $k_z$ is of the internal momentum
$\vec k=(\vec{k}_\bot, k_z)$, defined through
\begin{equation}
1-x = {e_1-k_z\over e_1 + e_2}, \qquad
x = {e_2+k_z \over e_1 + e_2},
\end{equation}
with $e_i = \sqrt{m_i^2 + \vec k^2}$. We then have
\be
M_0=e_1 + e_2,~~~~k_z = \,{xM_0\over 2}-{m_2^2+k_\perp^2 \over 2 xM_0},
\label{kz}
\en
and
\begin{equation}
        {{dk_z\over dx}} = \,{e_1 e_2\over x(1-x)M_0},
\end{equation}
which is the Jacobian of the transformation from
$(x, k_\bot)$ to $\vec k$.

\section{Numerical Results}
To examine numerically the form factor derived in Eq. (\ref{Hi}),
we need to specify the parameters appearing in $\phi_M(x,k_\bot)$.
To fit the meson masses,
in Ref.\cite{Ji} $m_u=0.22(0.25)$ GeV and 
$m_s=0.45(0.48)$ GeV are obtained with some interaction potentials,
while in Ref. \cite{Jaus} $m_u=0.25$ GeV and
$m_s=0.37$ GeV in the invariant meson mass scheme.
 Here we do not consider any potential form and scheme and
just use the decay constant $f_K=159.8~\rm {MeV}$
\cite{PDG00}, charge radius $\la r^2 \ra_K = 0.34~\rm {fm}^2$
\cite{Exp4}, and the quark masses of $m_{u,d}$ to constrain the
$s$ quark mass of $m_s$ and the scale parameter of
 $\omega$ in Eq. (\ref{gauss}). 
By using $m_u=m_d=250~\text{MeV}$ \cite{cheng}, we find that
$m_s=400~\text{MeV}$ and $\omega=0.38~\text{GeV}$. 
We note that the lower mass of $m_s$ should not affect the meson masses 
once we choose a suitable potential \cite{Ji} or scheme \cite{Jaus}.
Now, we use the
momentum distribution functions $\phi(x,k_\perp)_{\rm G}$
 to calculate the form factors $F(q_1^2,q^2_2)$ in time-like region of
$0 \leq q_1^2$ and $q_2^2 \leq M_K^2 \simeq 0.25~\mbox{GeV}^2$. In this
low energy region, we neglect the momentum dependence of the effective
vertex $C_W(q^2)$ in Eq. (\ref{Hi}), that is,
\be
C_W(q^2) \simeq C_W(0)
\label{appI}.
\en
We can use Eqs. (\ref{Hi}) and (\ref{appI})
to get the function $f(y)\equiv
 F(q_1^2,0)/F(0,0)$, where $y \equiv q_1^2/M^2_K$, and the
 result for $|f(y)|^2$ is shown in Fig.~3. From the
figure, we see that our result with the assumption of Eq. (\ref{appI})
agrees well with experimental data \cite{E799,E845,NA31}, especially
in the lower $y$ region. To get a better fit for a larger $y$,
we may use
\be
C_W(q^2) \simeq {C_W(0)\over (1-{q^2\over m^2_c})^n} \,.
\label{appII}
\en
As seen from Fig.~3, we find that the fit
for $n \, (<8)$ is better than that for $n-1$.
In particular, a larger value of $n$ is preferred if we disregard
the data from E845 at BNL \cite{E845} in Fig. 3. The experimental
result on $K_L\to\mu^+\mu^-\gamma$ from NA48 at CERN, which is
currently being analyzed \cite{NA48s}, should help to resolve this
matter.
To illustrate our results on the lepton pair decays, we shall take $n=0$
and $3$,
referring as (I) and (II), respectively.

The function of $f(y)$
is related to the differential decay rate of $K_L^0 \to l^+l^-\gamma$ by
\be
{d{\cal B}_{l^+l^-\gamma}\over dq^2_1\ \ }
\equiv{d\,\Gamma(K_L \to
l^+l^-\,\gamma)
\over{\Gamma(K_L \to \gamma\gamma)\,dq^2_1}}={2\over{q_1^2}}
\left({\alpha\over{3\pi}}\right)|f(y)|^2\,\lambda^{3/2}
\left(1,{q_1^2\over{M^2_{K_L}}},0\right)\,G_l(q_1^2),
\label{drate}
\en
where
\be
\lambda (a,b,c)=a^2+b^2+c^2-2(ab+bc+ca),
\en
and
\be
G_l(q^2)=\left(1-{4\,M^2_l\over{q^2}}\right)^{1/2}\left(1+{2\,M^2_l
\over{q^2}}\right).
\en
Integrating over $q^2_1$ in Eq. (\ref{drate}), we
get the branching ratios
\be
{\cal B}_{e^+e^-\gamma}&\equiv&{\Gamma(K_L^0\to e^+e^-\gamma)
\over{\Gamma(K_L^0\to\gamma \gamma)}}=1.64\,,\ 1.65 \times 10^{-2},
\non \\
{\cal B}_{\mu^+\mu^-\gamma}&\equiv&{\Gamma(K_L^0\to \mu^+\mu^-\gamma)
\over{\Gamma(K_L^0\to\gamma \gamma)}}=5.50\,,\
6.20 \times 10^{-4},
\en
for (I) and (II), respectively.
These values agree well with the experimental data:
${\cal B}_{e^+e^-\gamma}^{\text{exp}}
=(1.69 \pm 0.13)\times 10^{-2}$ \cite{PDG00} and ${\cal B}_{\mu^+\mu^
-\gamma}^{\text{exp}}=(6.11 \pm 0.31) \times 10^{-4}$ \cite{prl2m},
where we have used \cite{burk}
\be
\Gamma^{\text{exp}}(K_L^0\to\gamma \gamma)= [(5.92\pm 0.15)\times
10^{-4}]\Gamma^{\text{exp}}(K_L^0\to \text{all}).
\label{gg}
\en
On the other hand, our results are larger than ${\cal
B}_{e^+e^-\gamma}=1.59\times 10^{-2}$ and ${\cal
B}_{\mu^+\mu^-\gamma}=4.09 \times 10^{-4}$, respectively, obtained
in Ref. \cite{MT}, where the momentum dependence of the form
factor was neglected, i.e., $f(y)=1$.
 This inconsistency is reasonable because the
 kinematic factor $G_l(q^2)$ which leads the contribution
 at $q^2 \simeq 4 M^2_l$ is important, and the electron
 mass is very small so that $f(y)=1$ is only valid for
 the decay with an electron-positron pair. For the muonic pair
 case, since the mass of muon is not small, the effect
 of the deviation of neglecting the momentum dependence is
evident. This situation also occurs in the decays with two lepton pairs.

Next, Eq. (\ref{Hi}) can be also used to calculate the differential
 decay rates of $K_L \to l^+l^-l'^+l'^-$ by
\be
{d\,\Gamma(K_L \to l^+l^-l'^+l'^-)\over{\Gamma(K_L\to
\gamma\gamma)\,dq_1^2\,dq_2^2}}&=&{2\over{q_1^2q_2^2}}
\left({\alpha\over{3\pi}}\right)^2\left|{F(q_1^2,q_2^2)
\over{F(0,0)}}\right|^2\,\lambda^{3/2} \left(1,{q_1^2
\over{M^2_{K_L}}},{q_2^2\over{M^2_{K_L}}}\right)\,G_l(q_1^2)\,G_{l'}(q_2^2).
\en
After the integrations over $q^2_1$ and $q^2_2$, for (I) and (II)
we obtain the branching ratios as follows:
\be
{\cal B}_{e^+e^-e^+e^-}&\equiv&{\Gamma(K_L^0\to e^+e^-e^+e^-)
\over{\Gamma(K_L^0\to\gamma \gamma)}}= 6.61\,,\
6.74
\times 10^{-5},\non \\
{\cal B}_{\mu^+\mu^-e^+e^-}&\equiv&{\Gamma(K_L^0\to \mu^+\mu^-e^+e^-)
\over{\Gamma(K_L^0\to\gamma \gamma)}}=3.87
\,,\ 4.37
\times 10^{-6},\non \\
{\cal B}_{\mu^+\mu^-\mu^+\mu^-}&\equiv&{\Gamma(K_L^0\to
\mu^+\mu^-\mu^+\mu^-)\over{\Gamma(K_L^0\to\gamma \gamma)}}
=1.50
\,,\ 1.73
\times 10^{-9}.
\label{Bratios}
\en

In Table 1, we summary the experimental and theoretical values of the
decay branching ratios for the $K_L$ lepton pair modes.
The results of Ref. \cite{MT} correspond a point-like form factor, while
those in  Ref. \cite{ZG} are calculated at $O(p^6)$ in the
ChPT.

\begin{center}
{\small Table 1: Summary of the lepton pair decays of $K_L$.}
\end{center}
\begin{center}
\begin{tabular}{|c||c|c|c|c|c|c|} \hline
~Br~ & PDG \cite{PDG00} & new data &  (I) &  (II) & Ref. \cite{MT}
& Ref. \cite{ZG}
\\ \hline
$10^{2}\times{\cal B}_{e^+e^- \gamma}$
& $ 1.69\pm 0.09 $ & & $1.64 $ & $1.65$ & $1.59$  & $1.60\pm0.15$
\\ \hline
$10^{4}\times{\cal B}_{\mu^+\mu^- \gamma}$
&$5.49\pm0.49$ & $6.11\pm 0.31$ \cite{prl2m}
 & $5.50$
 & $6.20$
& $4.09$ & $4.01\pm 0.57$  \\ \hline
$10^{5}\times{\cal B}_{e^+ e^+ e^- e^-}$
& $6.93\pm0.20$ & $6.28\pm 0.65$ \cite{prl4e}
& $6.61$
& $6.74$
& $5.89$ &  $6.50$
\\
& & $6.20\pm0.69$ \cite{NA48} & &&&
\\ \hline
$10^{6}\times{\cal B}_{\mu^+\mu^-e^+e^-}$
&$4.9^{+11.3}_{-4.0}$
& $4.43\pm 0.84$ \cite{prl2e2m}
 & $3.87$
 & $4.37$
& $1.42$  &$2.20\pm 0.25$
\\ \hline
 $10^{9}\times{\cal B}_{\mu^+\mu^-\mu^+\mu^-}$ &  &
& $1.50$
& $1.73$
& $0.946$  &
$1.30\pm 0.15$ \\ \hline
\end{tabular}
\end{center}

 From Table 1, we may also combine the  experimental values
by assuming that they are uncorrelated and we find that
\be
{\cal B}_{K_L\to \mu^+\mu^-\gamma}^{\rm exp} &=&(5.93\pm0.26)\times 10^{-4}\,,
\nonumber\\
{\cal B}_{K_L\to e^+e^-e^+e^-}^{\rm exp}&=&(6.83\pm0.19)\times 10^{-5}\,,
\nonumber\\
{\cal B}_{K_L\to \mu^+\mu^-e^+e^-}^{\rm exp}&=& (4.44^{+0.84}_{-0.82})\times
10^{-6}\,.
\label{cexp}
\en
It is interesting to see that our results for $K_L\to
l^+l^-\gamma$ are larger than those in Refs. \cite{MT,ZG} and
agree very well with the experimental data. Furthermore, as shown
in  Eq. (\ref{Bratios}), those for $K_L\to e^+e^-e^+e^-$ and
$K_L\to \mu^+\mu^-e^+e^-$ also agree with
 the combined experimental values in Eq. (\ref{cexp}).
Here, we do not consider the interference effect \cite{MT,ZG}
 from the identical leptons in the final state. The reasons are
 given in the following.
When we use the non-point-like form factor, this effect is about
$0.5\%$ in the $e^+e^-e^+e^-$ mode \cite{ZG}, which is beyond
experimental access. For the  $\mu^+\mu^-\mu^+\mu^-$ mode, the
relative size of the interference
 effect is larger, but it is outside the scope of future experiments
because the total branching ratio is predicted to be
 about $8\times 10^{-13}$.

We now use the form factor $F(q_1^2,q_2^2)$ to calculate
the decays of $K_L \to l^+l^-$.
The decay branching ratios of the modes can be
generally decomposed in the following way
\be
{\cal B}_{l^+l^-}\equiv{\Gamma(K_L\to l^+l^-)\over{\Gamma(K_L
\to\gamma \gamma)}}= |\text{Im}\,\,{\cal A}_l|^2+|\text{Re}\,\,{\cal A}_l|^2,
\en
where $\text{Im}\,{\cal A}_l$ denotes the absorptive contribution and
$\text{Re}\,{\cal A}_l$ the dispersive one. The former can be determined
 in a model-independent form of
\be
|\text{Im}\,\,{\cal A}_l|^2={\alpha^2 M_l^2 \over{2\,M_{K_L}^2\beta_l}}
\,\Bigg[\text{ln}{1-\beta_l\over{1+\beta_l}}\Bigg]^2, \label{imaginary}
\en
where $\beta^2_l\equiv 1-4M^2_l/M^2_{K_L}$. The latter, however, can be
 rewritten as
the sum of SD and LD contributions,
\be
\text{Re}\,\,{\cal A}_l= \text{Re}\,\,{\cal A}_{l\,\text {SD}}+\text{Re}\,\,
{\cal A}_{l\,\text {LD}}.
\en
In the standard model, the SD part has been identified as the weak
contribution represented by one-loop $W$-box and $Z$-exchange diagrams
\cite{Geng,IL,SDmm},
while the LD
one is related to $F(q_1^2,q_2^2)$ by
\be
|\text{Re}\,\,{\cal A}_{l\,\text {LD}}|^2= {2\alpha^2M^2_l\beta_l
\over{\pi^2 M^2_{K_L}}}\,|\text{Re}\,\,{\cal R}_l(M^2_{K_L})|^2,
\en
where \cite{ABM}
\be
{\cal R}_l(P^2)={2i\over{\pi^2 M_K^2}}\int d^4 q\,{[P^2q^2-(P\cdot q)^2]
\over{q^2\,(P-q)^2\,[(q-p_l)^2-M^2_l]}}\, {F(q^2,(P-q)^2)\over{F(0,0)}}.
 \label{Rloop}
\en
 In general, an once-subtracted dispersion relation can be written for
$\rm{Re}\,{\cal R}$ as \cite{Berg}
\be
 \text{Re}\,\,{\cal R}_l(P^2)=\text{Re}\,\,{\cal
 R}_l(0)+{P^2\over{\pi}}\int^\infty_0 dP'^2{\text{Im}\,\,{\cal
R}_l(P'^2)\over{(P'^2-P^2)P'^2}}\,,
\label{ReR}
\en
where $\rm{Re}\,{\cal R}_l(0)$ can be obtained by applying Eq.
(\ref{Hi}) in the soft limit of $P \to 0$.

 For the $K_L \to e^+e^-$ decay, with $n=0$ and $3$ of (I) and (II)
in Eq. (\ref{appII})
we find that
\be
|\text{Re}\,\,{\cal A}_{e\,\text {LD}}|^2&=& 5.60\,,\ 6.52\times 10^{-9}\,,
\en
respectively. Since the SD part of $\rm{Re}\,{\cal A}_{e\,\text {SD}}$
can be neglected, we get
\be
{\cal B}_{e^+e^-}^I &=& 1.09 \times 10^{-8}\,,
\nonumber\\
{\cal B}_{e^+e^-}^{II} &=& 1.18 \times 10^{-8}\,,
\label{Bee}
\en
where we have used $|\rm{Im}\,{\cal A}_e|^2 = 5.32
 \times 10^{-9}$. In terms of the total decay branching ratio
 $B_{e^+e^-}=\Gamma (K_L\to e^+e^-)/\Gamma
(K_L\to all)$, the numbers in Eq. (\ref{Bee}) are about $6.5$ and
$7.0\times 10^{-12}$, respectively.
 Both results in Eq. (\ref{Bee}) are
consistent with the experimental value of
${\cal B}^{\rm {exp}}_{e^+e^-}=(1.5^{+1.0}_{-0.7}) \times 10^{-8}$
measured by E871 at BNL \cite{E871}, but they are
 lower than the value of $(1.52\pm0.09)\times 10^{-8}$
[$B_{e^+e^-}=(9.0\pm0.5)\times 10^{-12}$]
given by the calculation in Ref. \cite{Valencia} with the ChPT.
It is interesting to note that ${\cal B}_{e^+e^-}$ slowly increases as $n$
and  reaches $1.22\times 10^{-8}$ for $n=10$.
Clearly, our prediction is about $20\%$ smaller than that in the
ChPT \cite{Valencia}.

 For the $K_L \to \mu^+\mu^-$ decay,
by subtracting between the value of $|\rm{Im}\,{\cal A}_\mu|^2 = 1.20
\times 10^{-5}$ from the experimental data of ${\cal B}^{\rm
{exp}}_{\mu^+\mu^-}=(1.21\pm 0.04) \times 10^{-5}$
\cite{PDG00,E871mm}, we obtain that
\be
|\text{Re}\,\,{\cal A}_\mu|^2 \leq 7.2 \times 10^{-7}\ \ \ (90\%\ C.L.).
\label{Limit}
\en
In the standard model, we have that \cite{Kmm2,Buras1}
\be
|\text{Re}\,\,{\cal A}_{\mu\,\text {SD}}|^2{\cal B}_{K_L\to\gamma\gamma}
&=& 0.9\times 10^{-9}(1.2-\bar{\rho})^2\left[{\bar{m}_t(m_t)\over 170\
\rm{GeV}}\right]^{3.1}\left[{|V_{cb}|\over 0.040}\right]^4\,,
\label{SMf}
\en
where $\bar{\rho}=\rho(1-\lambda^2/2)$.
Using the parameters of $\bar{m}_t(m_t)=166\ \rm{GeV}$,
$|V_{cb}|=0.041$ and $\bar{\rho}\simeq 0.224$ \cite{SDmm,CKMfit},
from Eqs. (\ref{gg}) and (\ref{SMf}) we get
 \be
\text{Re}\,\,{\cal A}_{\mu\,\text {SD}}\simeq -1.22 \times 10^{-3}\,,
\label{SMA}
\en
which  is larger than the limit in Eq. (\ref{Limit}). It is clear
that the value of $\rm{Re}\,{\cal A}_{\mu\,\text {LD}}$ has to be
either very small for the same sign as $\rm{Re}\,{\cal
A}_{\mu\,\text {SD}}$ or the same order but the opposite sign.

For the case of (I), from
 Eq. (\ref{ReR}) we find
\be
\text{Re}\,\,{\cal A}^I_{\mu\,\text {LD}}=-1.11\times
10^{-3}\,,
\label{LDAI}
\en
which is very close to the SD value in Eq. (\ref{SMA}) and
clearly ruled out if the absolute sign in Eqs. (\ref{SMA}) and
(\ref{LDAI})
are the same. However, if the relative sign is opposite, the limit in
Eq. (\ref{Limit}) can be satisfied for certain values of $\rho$.
 From Eqs. (\ref{SMf}),
(\ref{Limit}), and (\ref{LDAI}), by taking $\bar{m}_t(m_t)=166\
\rm{GeV}$ and $|V_{cb}|=0.041$ we extract that
\be
\bar{\rho} >-0.37\ \ {\rm or}\ \ \rho>-0.38\ \ (90\%\,C.L.)\,.
\label{rhoI}
\en
We note that the limit in Eq. (\ref{rhoI}) is close to that
in Eq. (41) of Ref. \cite{Kmm2}.
This result is not surprising.  If we fit $F(q_1^2,q_2^2)$ in Eq.
(\ref{Hi}) with Eq. (14) of Ref. \cite{Kmm2} given by
\be
f(q_1^2,q_2^2)={F(q_1^2,q_2^2)\over F(0,0)}=
1+\alpha\left({q_1^2\over q_1^2-m_\rho^2}+{q_2^2\over
q_2^2-m_\rho^2}\right)+
\beta{q_1^2q_2^2\over (q_1^2-m_\rho^2)(q_2^2-m_\rho^2)}\,,
\en
we find that $\alpha \simeq -0.585$ and $\beta \simeq 0.191$ and
thus
\be
1+2\alpha+\beta=2.16 \times 10^{-2}\simeq 0\,,
\en
which satisfies the bound of Eq. (35) in Ref. \cite{Kmm2}.
Similarly, for (II) we obtain
\be
\text{Re}\,\,{\cal A}^{II}_{\mu\,\text {LD}}=-1.38
\times 10^{-4}\,.
\label{LDAII}
\en
It is very interesting to see that the value in Eq. (\ref{LDAII})
is much smaller than $\text{Re}\,{\cal A}_{\mu SD}$ in Eq.
(\ref{SMA}), which is exactly the case discussed in Ref.
\cite{Geng}. From Eq. (\ref{LDAII}), with the same parameters as (I), 
we find that
\be
\bar{\rho} >0.63,~0.41~~~{\rm or}~~~\rho >0.65,~0.42~~~ (90\%\,C.L.)\, \label{rhoII}
\en
for the same and opposite signs between $\text{Re}\,{\cal A}_{\mu
SD}$ and $\text{Re}\,{\cal A}^{II}_{\mu\,\text{LD}}$,
respectively.
We note that the limits in Eq. (\ref{rhoII}) do not agree with the
recent global fitted value of $\bar{\rho}=0.224\pm0.038$
\cite{SDmm,CKMfit}, which may not be unexpected since (i) we have
not included various possible ranges of $\bar{m}_t(m_t)$,
$|V_{cb}|$, and quark masses in the calculation and (ii) we still
need to fix $n$ in Eq. (\ref{appII}) and modify the form of
$C_W(q^2)$ \cite{GH}. However, the important message here is that
the LD dispersive contribution in $K_L\to\mu^+\mu^-$ is calculable
in the LFQA. From our preliminary results, it seems that
$\text{Re}{\cal A}_{\mu LD}$ is indeed small as anticipated many
years ago in Ref. \cite{Geng}. Moreover, our approach here
provides another useful tool for the decays beside the ChPT.

\section{Conclusions}

In this work, we have studied the $K_L$ lepton pair decays of
 $K_L \to l^+\l^-\gamma$ and $K_L \to l^+l^-l'^+l'^-$
in the light-front QCD framework. In our calculations, we have
adopted the Gaussian-type wave function and assumed the form of
the effective vertex $C_W(q^2)$ in Eq. (\ref{appII}) to account
for the  momentum dependences  in the low energy region. We have
calculated the relative form factors of the leptonic decays vs. the two-photon decay, and have showed that our results on the decay branching ratios of $K_L\to
l^+l^-\gamma$ and $e^+e^-l^+l^-\ (l=e,\mu)$ agree well with the
experimental data. The remarkable agreements indicate that our
form for $C_W(q^2)$
  is quite reasonable, but the number of $n$
 still needs to be fixed.
Furthermore, all our predicted values for these decays are larger
than those in the ChPT \cite{MT,ZG}, in particular for the modes
of $\mu^+\mu^-\gamma$ and $\mu^+\mu^- e^+e^-$ for which the
$O(p^6)$ ChPT results in Ref. \cite{ZG} are ruled out by the new
experimental data \cite{prl2m,prl2e2m}.
 On the other hand, for $K_L\to e^+e^-$, we have found that ${\cal
B}_{e^+e^-}$ is between $1.09$ and $1.22\times 10^{-8}$
  for
$n=(0,10)$, which
are lower than $(1.52\pm 0.09)\times 10^{-8}$ in the ChPT
\cite{Valencia}. For $K_L\to
\mu^+\mu^-$, we have demonstrated that the long-distance
dispersive contribution is possibly small. However, to get  a
meaningful constraint on the CKM parameters, further theoretical
studies \cite{GH} as well as more precise experimental data such
as those from NA48 at CERN \cite{NA48s} on the spectra of the pair
decays are needed.
Finally, we remark that our approach cannot calculate the absolute decay
widths of $K_L\to l^+l^-\gamma$ and $K_L\to \gamma\gamma$.


\acknowledgments
We would like to thank K. Terasaki for useful data. This work was supported
in part by the National Science Council of R.O.C. under the Grant
No. NSC90-2112-M-007-040.


\newpage
\parindent=0 cm
\centerline{\bf FIGURE CAPTIONS}
\vskip 0.5 true cm

{\bf Fig. 1 } Feynman diagrams for the meson (a) decay constant
and (b) normalization. 

\vskip 0.25 true cm 
{\bf Fig. 2 }  Feynman triangle diagrams with (a) and (b) corresponding to the LF valence configuration. Empty circles indicate LF wave functions. 

\vskip 0.25 true cm 
{\bf Fig. 3} The $y$-dependent behavior of
$|f(y)|^2$, where the lines from bottom to top corresponding to
$n=0,1,\cdots,10$ are obtained by this work with $f_K=159.8
\rm{MeV}$ and $m_s=400 \rm {MeV}$ and the experimental data are
taken from E799 at FNAL \cite{E799}, E845 at BNL \cite{E845}, and NA31 at 
CERN \cite{NA31}, respectively.

\newpage

\begin{figure}[h]
\includegraphics{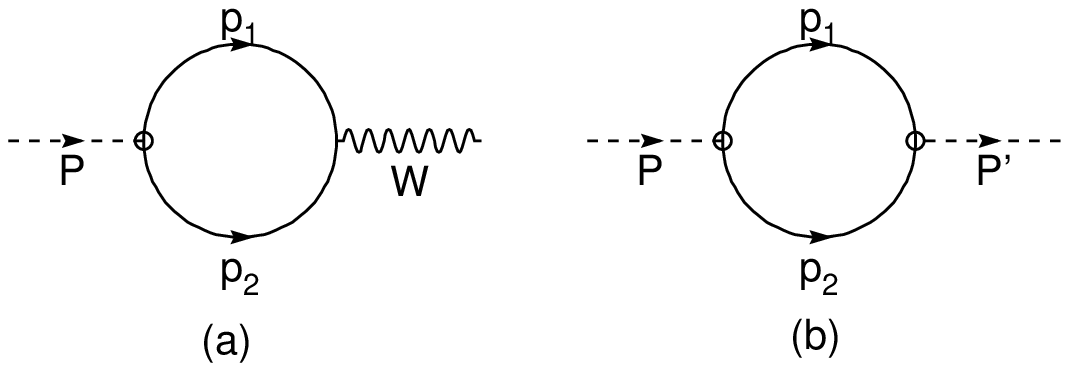}
\caption{}
\vskip 13.cm
\end{figure}

\newpage
\begin{figure}[h]
\includegraphics{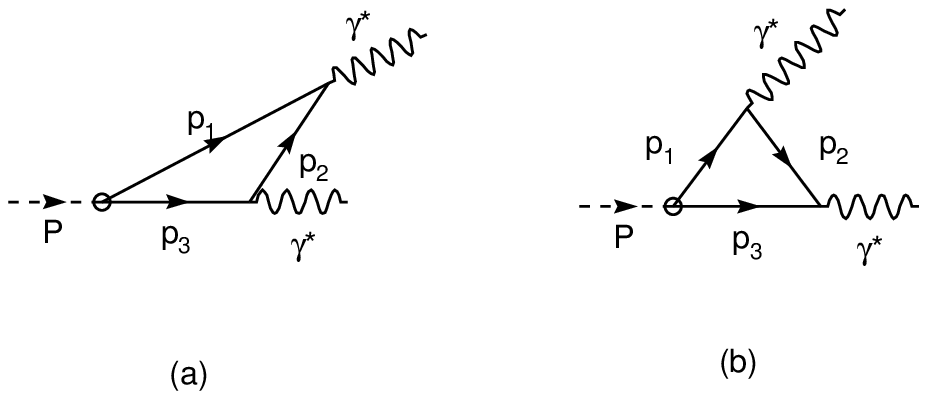}
\vskip 23.cm
\caption{}
\end{figure}

\newpage
\begin{figure}[h]
\includegraphics{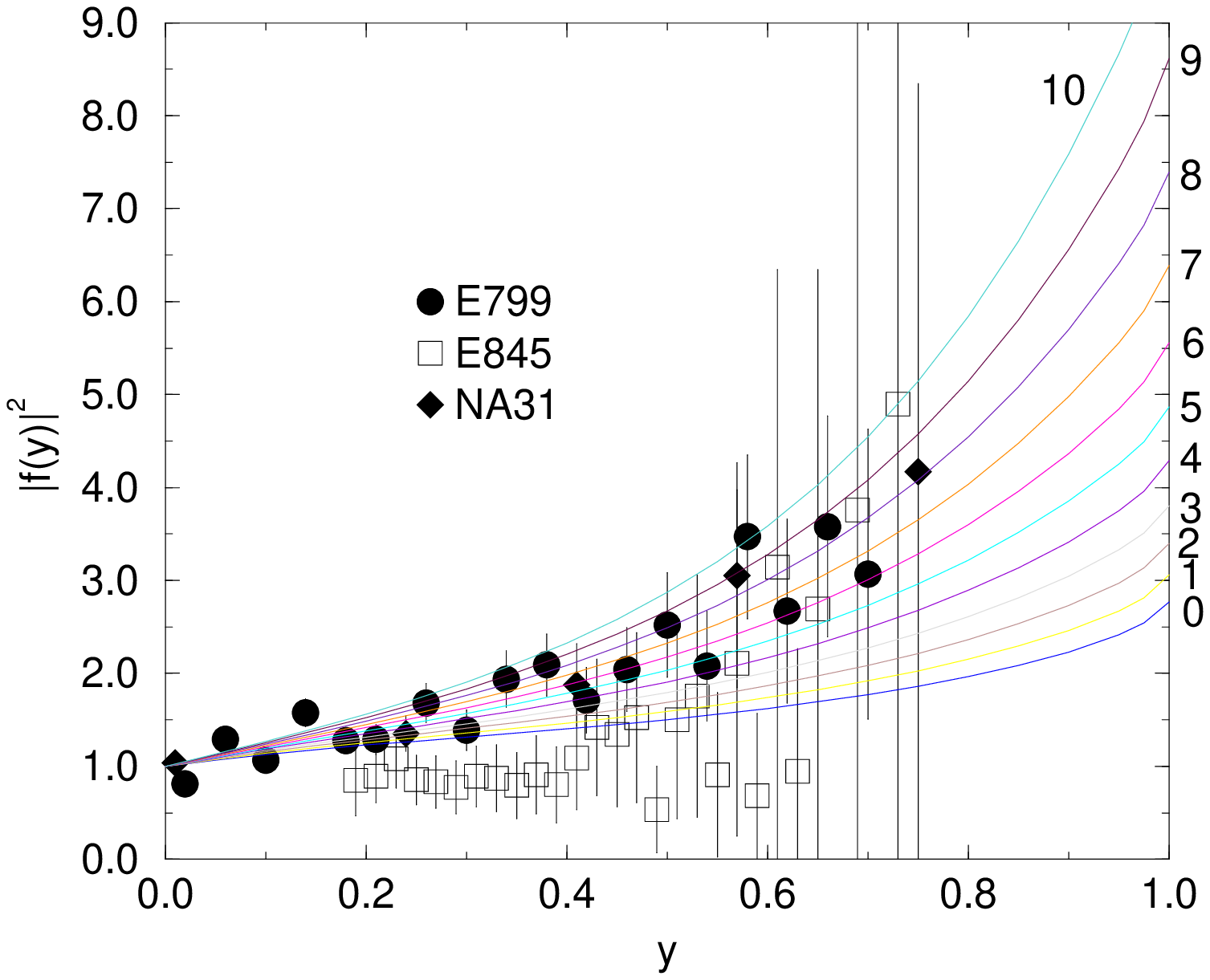} \vskip 23.cm \caption{}
\end{figure}

\end{document}